\documentclass[nofootinbib,amsmath,notitlepage,preprintnumbers]{revtex4-1}
\usepackage{multirow}
\usepackage{amssymb,esvect,amsmath,graphicx,latexsym,amsthm,slashed,eso-pic}
\usepackage{units}
\usepackage{hyperref}
   \DeclareMathOperator{\gev}{GeV}       \DeclareMathOperator{\g}{g}        
         \newcommand{\cO}{{\cal O}}

\newcommand{\pL}{\left(} \newcommand{\pR}{\right)} \newcommand{\bL}{\left[} \newcommand{\bR}{\right]}    
\newcommand{\beq}{\begin{equation}} \newcommand{\eeq}{\end{equation}}
\newcommand{\bea}{\begin{eqnarray}} \newcommand{\eea}{\end{eqnarray}}

\newcommand{\lsim}{\mathrel{\hbox{\rlap{\lower.55ex\hbox{$\sim$}} \kern-.3em \raise.4ex \hbox{$<$}}}}
\newcommand{\gsim}{\mathrel{\hbox{\rlap{\lower.55ex\hbox{$\sim$}} \kern-.3em \raise.4ex \hbox{$>$}}}}

\def\lsim{\mathrel{\raise.3ex\hbox{$<$\kern-.75em\lower1ex\hbox{$\sim$}}}}
\def\gsim{\mathrel{\raise.3ex\hbox{$>$\kern-.75em\lower1ex\hbox{$\sim$}}}}

\newcommand{\be}{\begin{eqnarray}}
\newcommand{\ee}{\end{eqnarray}}

\newcommand{\benum}{\begin{enumerate}}
\newcommand{\eenum}{\end{enumerate}}
\newcommand{\bi}{\begin{itemize}}
\newcommand{\ei}{\end{itemize}}

\begin{document}

\preprint{FERMILAB-PUB-19-186-A
}

\title{Dark Radiation and Superheavy Dark Matter from Black Hole Domination}

\author{Dan Hooper$^{a,b,c}$}
\thanks{ORCID: http://orcid.org/0000-0001-8837-4127}

\author{Gordan Krnjaic$^{a}$}
\thanks{ORCID: http://orcid.org/0000-0001-7420-9577}

\author{Samuel D.~McDermott$^{a}$}
\thanks{ORCID: http://orcid.org/0000-0001-5513-1938}

\affiliation{$^a$Fermi National Accelerator Laboratory, Theoretical Astrophysics Group, Batavia, IL 60510}
\affiliation{$^b$University of Chicago, Kavli Institute for Cosmological Physics, Chicago, IL 60637}
\affiliation{$^c$University of Chicago, Department of Astronomy and Astrophysics, Chicago, IL 60637}

\date{\today}

\begin{abstract}

If even a relatively small number of black holes were created in the early universe, they will constitute an increasingly large fraction of the total energy density as space expands. It is thus well-motivated to consider scenarios in which the early universe included an era in which primordial black holes dominated the total energy density. Within this context, we consider Hawking radiation as a mechanism to produce both dark radiation and dark matter. If the early universe included a black hole dominated era, we find that Hawking radiation will produce dark radiation at a level $\Delta N_{\rm eff} \sim 0.03-0.2$ for each light and decoupled species of spin 0, 1/2, or 1. This range is well suited to relax the tension between late and early-time Hubble determinations, and is within the reach of upcoming CMB experiments. The dark matter could also originate as Hawking radiation in a black hole dominated early universe, although such dark matter candidates must be very heavy ($m_{\rm DM} \gsim 10^{11} \, {\rm GeV}$) if they are to avoid exceeding the measured abundance.

\end{abstract}

\maketitle

\section{Introduction}

It has long been appreciated that inhomogeneties in the early universe could lead to the formation of primordial black holes~\cite{Carr:1974nx}. In particular, inflationary models that predict a significant degree of non-Gaussianity could result in a cosmologically relevant abundance of black holes, typically with masses near the value enclosed by the horizon at or near the end of inflation, 
\be
M_{\rm hor} \sim \frac{M_{\rm Pl}^2}{2H_I} \sim 10^4 \, {\rm g}  \left(\frac{ 10^{10} \, {\rm GeV} }{H_I}\right)~,~~
\ee
 where $M_{\rm Pl} = 1.22 \times 10^{19}$ GeV is the Planck mass and $H_I$ is the Hubble rate at the time of black hole formation~\cite{GarciaBellido:1996qt,Kawasaki:2016pql,Clesse:2016vqa,Kannike:2017bxn,Kawasaki:1997ju,Cai:2018rqf,Yoo:2018esr,Young:2015kda,Clesse:2015wea,Hsu:1990fg,La:1989za,La:1989st,La:1989pn,Weinberg:1989mp,Steinhardt:1990zx,Accetta:1989cr,Holman:1990wq,Hawking:1982ga,Khlopov:1980mg}. Alternatively, phase transitions in the early universe may have provided the conditions under which substantial quantities of black holes could have formed~\cite{Sobrinho:2016fay,Rubin:2000dq,Jedamzik:1999am,Byrnes:2018clq}. 

Black holes with initial masses below $M_i \lsim 5 \times 10^{8}$ grams will disappear through the process of Hawking evaporation prior to Big Bang Nucleosynthesis (BBN), and are thus almost entirely unconstrained by existing observations. Furthermore, black holes evolve like matter in the early universe, constituting an increasingly large fraction of the total energy density as the universe expands (up to the time of their evaporation). From this perspective, it would not be surprising if the early universe included an era in which primordial black holes made up a substantial fraction or even dominated the total energy density.

Unlike most other mechanisms for particle production, the process of Hawking evaporation generates particles democratically, producing all particle species regardless of their assigned charges or couplings. If there exist any stable particles without significant couplings to the Standard Model (SM), they would be produced through Hawking evaporation without subsequently thermalizing, but still contributing to the universe's total energy density. This makes the evaporation of primordial black holes a particularly attractive mechanism for the production of both dark radiation and dark matter \cite{Fujita:2014hha,Lennon:2017tqq,Morrison:2018xla}. 

A $4.4\sigma$ discrepancy has been reported between the value of the Hubble constant as determined from local measurements~\cite{Riess:2019cxk,Riess:2018byc,Riess:2016jrr} and as inferred from the temperature anisotropies of the cosmic microwave background (CMB)~\cite{Aghanim:2018eyx}. This tension can be substantially relaxed if there exist one or more exotic particle species that contribute to the relativistic energy density of the universe in the period leading up to matter-radiation equality, as often parameterized in terms of the quantity, $\Delta N_{\rm eff}$~\cite{Bernal:2016gxb,Aylor:2018drw,Weinberg:2013kea,Shakya:2016oxf,Berlin:2018ztp,DEramo:2018vss,Dessert:2018khu,Escudero:2019gzq} (for alternative explanations, see Refs~\cite{Poulin:2018cxd,Poulin:2018dzj,Poulin:2018zxs,Agrawal:2019lmo,Bringmann:2018jpr,Kreisch:2019yzn}). If there exist any light and long-lived decoupled particles (such as axions, for example), then any black holes that are present in the early universe will produce a background of such states, contributing to the value of $\Delta N_{\rm eff}$. If the early universe included an era in which the energy density was dominated by black holes, we find that Hawking radiation will contribute to the dark radiation at a level that can naturally address the tension between late and early-time Hubble determinations, $\Delta N_{\rm eff} \sim 0.03-0.2$ for each light and decoupled species of spin 0, 1/2, or 1. 

The null results of direct detection~\cite{Aprile:2018dbl,Akerib:2016vxi,Cui:2017nnn} and collider searches for dark matter provide us with motivation to consider dark matter candidates that were never in thermal equilibrium in the early universe, but that were instead produced through other mechanisms, such as misalignment production~\cite{Kolb:1990vq,Kim:1986ax,Turner:1989vc}, out-of-equilibrium decays~\cite{Gelmini:2006pq,Gelmini:2006pw,Merle:2015oja,Merle:2013wta,Kane:2015jia}, or gravitational production during inflation~\cite{Chung:1998ua,Chung:1998rq,Chung:2001cb}. Hawking evaporation is a theoretically well motivated way to generate dark matter particles that would not lead to observable signals in existing experiments. As we will show, very heavy dark matter candidates ($m_{\rm DM} \gsim 10^9$ GeV) can naturally be generated with the measured abundance in scenarios in which the early universe included a black hole dominated era.

In this paper, we revisit primordial black holes in the early universe, focusing on scenarios which include a black hole dominated era prior to BBN. In such scenarios, we find that the products of Hawking evaporation can naturally contribute substantially to the abundance of dark radiation, at a level of $\Delta N_{\rm eff} \sim 0.03-0.2$ for each light and decoupled species. An abundance of dark radiation in this range would help to relax the reported Hubble tension, and is projected to be within the reach of upcoming stage IV CMB experiments. We also consider the production of dark matter through Hawking radiation in a black hole dominated early universe, finding that the measured dark matter abundance can be easily accommodated if the mass of the dark matter candidate lies in the range between $m_{\rm DM}\sim 10^{9}$ GeV and the Planck scale.

\section{Black Holes in the Early Universe}

\subsection{Evaporation Preliminaries}

A black hole with mass $M_{\rm BH}$ loses mass through the process of Hawking evaporation~\cite{Hawking:1974sw}, at a rate given by: 
\begin{eqnarray}
\frac{dM_{\rm BH}}{dt} = -\frac{\mathcal G \, g_{\star,H}(T_{\rm BH}) \, M_{\rm Pl}^4}{30720 \, \pi \,  M_{\rm BH}^2} 
\simeq -7.6\times 10^{24} \, {\rm g \, s^{-1}} ~ g_{\star, H}(T_{\rm BH})      \left(               \frac{\rm g}{      M_{\rm BH }  } \right)^2, 
\label{rate}
\end{eqnarray}
where $\mathcal G \approx 3.8$ is the appropriate graybody factor, the temperature of a black hole is
\be
T_{\rm BH} = \frac{ M^2_{\rm Pl} }{8 \pi M_{\rm BH}} \simeq 
1.05\times 10^{13} \,  {\rm GeV}  \left(               \frac{\rm g}{      M_{\rm BH }  } \right),
\ee
and $g_{\star, H}(T_{\rm BH})$ counts {\it all existing} particle species with masses below $T_{\rm BH}$ ~\cite{1990PhRvD..41.3052M,1991PhRvD..44..376M} 
according to the prescription 
\be
\label{ghdof}
\hspace{-0.5cm
}g_{\star,H}(T_{\rm BH}) \equiv \sum_i w_i g_{i,H}   ~~~~~,~~~~~ g_{i,H}   = 
\begin{cases}
1.82 ~& s = 0 \\
1.0 ~& s = \nicefrac{1}{2} \\
0.41~& s = 1 \\
0.05~ & s = 2
\end{cases}~,
\ee
where $w_i = 2s_i+1$ for massive particles of spin $s_i$, $w_i= 2$ for massless particles with $s_i > 0$, and $w_i=1$ for $s_i =0$ species.
At BH temperatures well above the electroweak scale BH evaporation emits the full SM particle spectrum according to their $g_{\star, H}$ weights; at temperatures below the MeV scale, only photons and neutrinos are emitted, so in these limits we have
\be
 g_{\star, H}(T_{\rm BH})   \simeq 
\begin{cases}
108~,~&T_{\rm BH} \gg {\rm 100 \, GeV} ~, ~M_{\rm BH} \ll 10^{11} {\rm g}\\
~~7~~,~&T_{\rm BH} \ll ~~{\rm MeV}~,~~~~ M_{\rm BH} \gg 10^{16} {\rm g}~~.
\end{cases}
\ee
Assuming $g_{\star, H}(T_{\rm BH}) $ is approximately constant (which is always true for BH that evaporate entirely
to SM radiation before BBN), integrating Eq.~(\ref{rate}) yields the time dependence of a BH with initial mass $M_i$
\begin{equation}
M_{\rm BH}(t) = M_i \bigg(1-\frac{t}{\tau}\bigg)^{1/3},
\end{equation}
and its evaporation time $\tau$ can be written as
\begin{eqnarray}
\label{tevap}
\tau \approx 1.3\times 10^{-25} \, {\rm s\, g^{-3}} \int^{M_i}_0 \frac{dM_{\rm BH} M_{\rm BH}^2}{g_{\star, H}(T_{\rm BH})} 
\approx 4.0 \times 10^{-4} \, {\rm s} ~ \bigg(\frac{M_i}{10^8 \, {\rm g}}\bigg)^3 \bigg(\frac{108}{ g_{\star, H}(T_{\rm BH})}\bigg). \nonumber
\end{eqnarray}
Although black holes can undergo mergers to form larger black holes and gain mass through accretion in the early universe, we expect these processes to play an important role only at very early times, corresponding to $T \gsim 10^{8}$ GeV $\times (10^8 \, {\rm g}/M_{\rm BH})^{3/4}$ (see Appendices~\ref{mergers} and \ref{accretion}). With this in mind, one can think of the quantity, $M_i$, as the mass of a given black hole after these processes have ceased to be efficient.

\subsection{Radiation Dominated Early Universe}
\label{radiation-domination}
Assuming for the moment that the universe was radiation dominated throughout its early history, the expansion rate is given by:
\begin{equation}
H^2 \equiv \bigg(\frac{\dot{a}}{a}\bigg)^2 = \frac{8 \pi G \rho_{R}}{3} = \frac{1}{4 t^2}~~~,~~ ~~
\rho_R(T) = \frac{\pi^2 g_\star(T)}{30} T^4~,~~~
\end{equation}
where $G = M_{\rm Pl}^{-2}$, $\rho_R$ is the energy density in radiation,
 and $g_{\star}(T)$ is the effective number of relativistic SM degrees-of-freedom 
with equilibrium distributions
\be
\label{gstar}
g_{\star}(T) = \sum_{B} g_B \left( \frac{T_B}{T} \right)^4 +  \frac{7}{8}\sum_{F} g_F \left( \frac{T_F}{T} \right)^4~,
\ee
evaluated at the photon temperature $T$, where the sum is over all bosons/fermions $B/F$ with temperatures $T_{B/F}$ and spin states  $g_{B/F}$. At 
the time of evaporation $t = \tau$, the Hubble rate is $H = 1/2 \tau$, which corresponds to a SM temperature of   
\begin{equation}
T_\tau \simeq 40 \, {\rm MeV}  \bigg(\frac{10^8\,{\rm g}}{M_i}\bigg)^{3/2} \, \bigg(\frac{g_{\star, H}(T_{\rm BH})}{108}\bigg)^{1/2} \, \bigg(\frac{14}{g_{\star}(T_{\tau})}\bigg)^{1/4}.
\end{equation}

As the universe expands, any population of primordial black holes that is present will evolve to constitute an increasingly large fraction of the universe's total energy density, with the ratio $\rho_{\rm BH}/\rho_{R}$ growing proportionally with the scale factor. For very massive and long-lived black holes, this will continue until the epoch of matter-radiation equality at which point the black holes will constitute some or all of the dark matter (see, for example, Refs~\cite{Bird:2016dcv,Carr:2016drx}). If the black holes are relatively light, however, they will evaporate long before this point in cosmic history.

A universe that contains both radiation and a population of black holes will evolve as
\begin{equation}
H^2 \equiv \bigg(\frac{\dot{a}}{a}\bigg)^2 = \frac{8\pi G}{3} \bigg(    \frac{     \rho_{{R},i}  }{  a^{4} }  + \frac{  \rho_{{\rm BH},i} }{a^{3}} \bigg),
\end{equation}
where $\rho_{{R},i}$ and $\rho_{{\rm BH},i}$ denote the energy densities in radiation and in black holes at some initial time, respectively. If the universe starts out dominated by radiation with a temperature, $T_i$, the fraction of the energy density that consists of black holes will grow by a factor of $\sim (T_i/40 \, {\rm MeV}) (M_i/10^8\,{\rm g})^{3/2}$ by the time that they evaporate. Thus, if the initial BH density faction $f_i$ satisfies 
\be
\label{domination-fraction}
\hspace{-3cm}\text{(Eventual BH Domination)}~~~~~~~~f_i \equiv \frac{ \rho_{{\rm BH,}i}  }{  \rho_{{R},i} } \gtrsim 
4 \times 10^{-12} \left(\frac{ 10^{10} \, {\rm GeV}  }{ T_i} \right)\left( \frac{ 10^8\,{\rm g} }{ M_i }\right)^{3/2},
\ee
the black hole population will ultimately come to dominate the energy density of the early universe. In this sense, black hole domination is at attractor solution to the evolution of the early universe. From this perspective, it would not be surprising if the early universe included a black hole dominated era.

\subsection{Black Hole Dominated Early Universe}
In light of Sec.\ref{radiation-domination}, it is generic to expect a period of BH domination even if the initial BH energy density is a 
small fraction of the initial radiation density, as shown in Eq.~(\ref{domination-fraction}). If this condition is satisfied in the early universe, 
Hawking radiation plays a key role in reheating the universe and setting the initial conditions for all subsequent cosmological epochs,
including Big Bang Nucleosynthesis (BBN). Thus, in this section and for the remainder of this work, we will assume that the early universe is black hole
dominated at early times and the subsequent era of radiation domination, which contains BBN, arises entirely from Hawking radiation.

Once the black holes have entirely evaporated (at $t = \tau$), the universe will be filled with the products of Hawking radiation. In the case of SM evaporation products, these particles will thermalize to form a bath with the following temperature:
\begin{equation}
T_{\rm RH} \simeq 50 \, {\rm MeV}\,  \bigg(\frac{10^8\,{\rm g}}{M_i}\bigg)^{3/2} \, \bigg(\frac{g_{\star, H}(T_{\rm BH})}{108}\bigg)^{1/2} \, \bigg(\frac{14}{g_{\star}(T_{\rm RH})}\bigg)^{1/4}.
\end{equation}
To preserve the successful predictions of  BBN, we will limit ourselves to the case in which $T_{\rm RH} \gsim 3$ MeV, corresponding to black holes lighter than $M_i \lsim 6\times 10^8$ g. For this range of masses, $T_{\rm BH} \gg 10$ TeV through evaporation, so the SM contribution to $g_{\star, H}$ is always $\simeq$\,108. One might na\"ively  be concerned that scenarios with low reheating temperatures would pose a challenge for baryogenesis, as in standard cosmology this requires the existence of light particles that violate baryon number. The Hawking evaporation of light black holes, however, populates the universe with particles that are much heavier than the temperature of the SM bath, and thus provides a natural means by which to generate the universe's baryon asymmetry~\cite{Hawking:1974rv,Carr:1976zz,Zeldovich:1976vw,Toussaint:1978br,Grillo:1980rt,Turner:1979bt,Hawking:1982ga,Barrow:1990he,Majumdar:1995yr,Polnarev:1986bi,Bugaev:2001xr,Upadhyay:1999vk,Kuzmin:1985mm,Fukugita:1986hr,Harvey:1990qw,Baumann:2007yr,Morrison:2018xla,Hamada:2016jnq,Hook:2014mla,Fujita:2014hha}.

The Hawking evaporation of primordial black holes will produce all particle species, including any that exist beyond the limits of the SM. Unlike most other mechanisms for particle production in the early universe, Hawking radiation is universal across all species, depending only on the mass and spin of the radiated degrees-of-freedom. Any products of Hawking evaporation that possess significant couplings to the SM will rapidly thermalize with the surrounding bath. On the other hand, Hawking evaporation could also produce particle species with extremely feeble (perhaps only gravitational) couplings to the SM, which would not thermalize. If relatively light, such particles would constitute dark radiation, and would be observable through their contribution to the effective number of neutrino species, $N_{\rm eff}$. Heavier evaporation products, on the other hand, would quickly become non-relativistic and could plausibly contribute to or constitute the dark matter of the universe.

\section{Dark Hawking Radiation and the Contribution to $\Delta N_{\rm eff}$}

As the universe expands, the energy density in black holes evolves as follows:
\begin{equation}
\frac{d\rho_{\rm BH}}{dt} = -3 \rho_{\rm BH} H + \rho_{\rm BH} \frac{dM_{\rm BH}}{dt}\frac{1}{M_{\rm BH}},
\end{equation}
where the first term corresponds to dilution from Hubble expansion, while the second term is from Hawking evaporation as described in Eq.~(\ref{rate}). In concert, the energy density in radiation evolves as follows:
\begin{eqnarray}
\frac{d\rho_{R}}{dt} = -4 \rho_{R} H - \rho_{\rm BH} \frac{dM_{\rm BH}}{dt}\bigg|_{\rm SM} \frac{1}{M_{\rm BH}},
\end{eqnarray}
where the second term is the result of energy being injected into the universe from the evaporating black holes.

\begin{figure}[t]
\hspace{-0.3cm}
\includegraphics[scale=0.4]{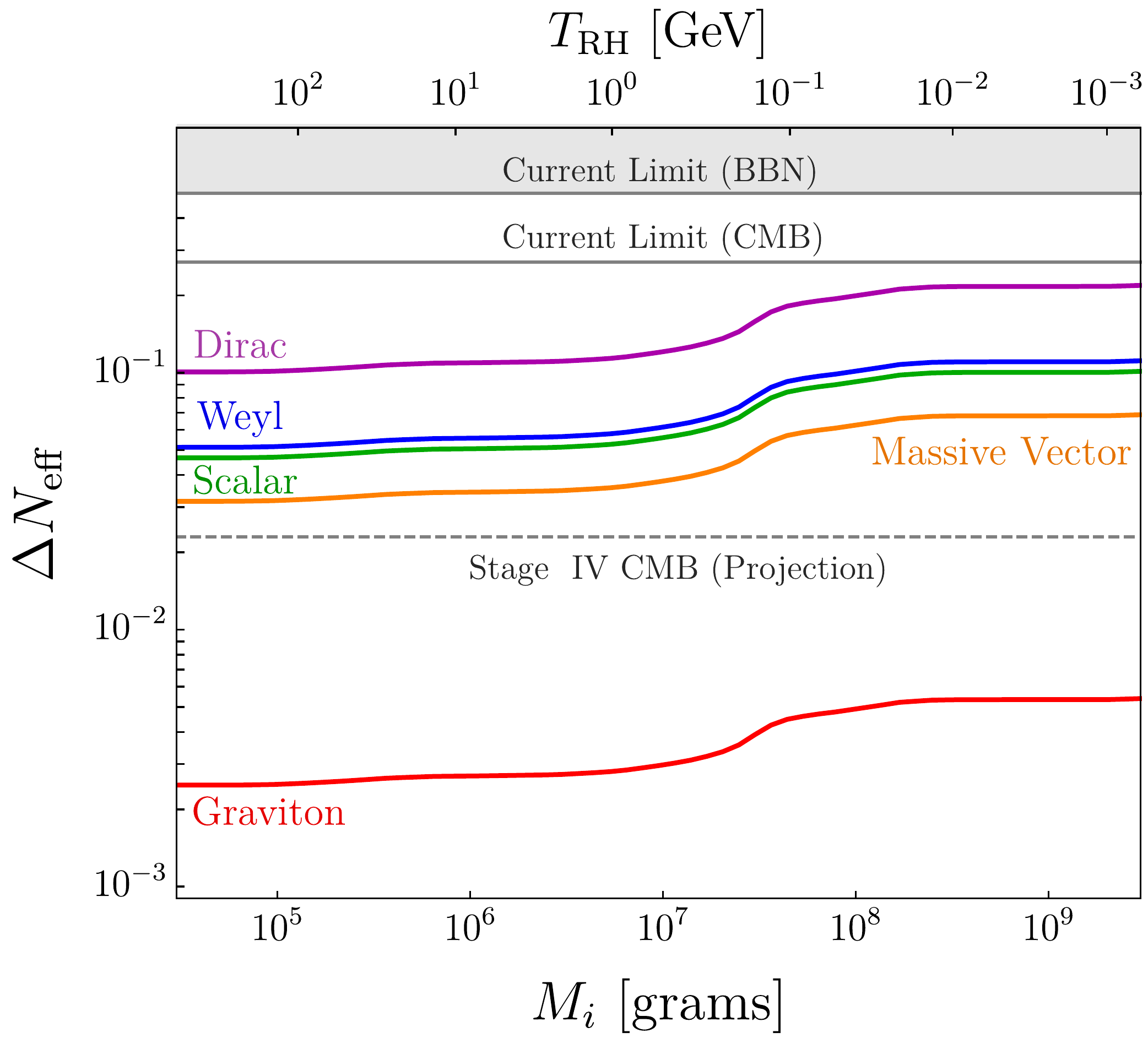}~~~~
\includegraphics[scale=0.4]{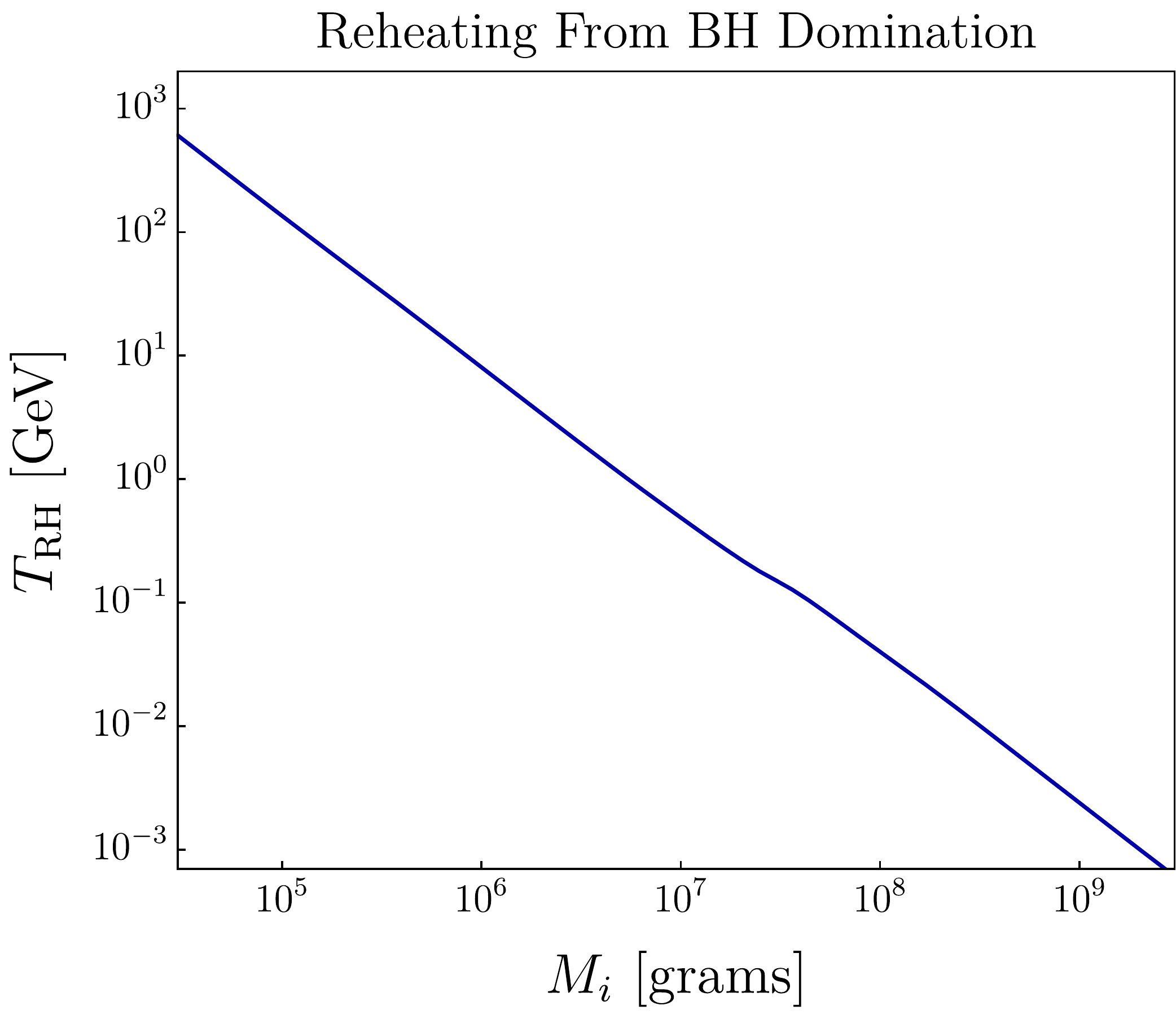}
\caption{In the left frame, we show the contribution to the effective number of neutrino species from Hawking evaporation in a scenario in which the early universe included a black hole dominated era. Results are shown for a single, light decoupled state that is either a Dirac fermion, Weyl Fermion, real scalar, massive vector, or massless spin-2 graviton, as a function of the temperature of the universe after black hole evaporation. Also shown are the current constraints~\cite{Aghanim:2018eyx} as well as the projected sensitivity of stage IV CMB measurements~\cite{Abazajian:2016yjj,Baumann:2017gkg,Hanany:2019lle}. For $\Delta N_{\rm eff} \sim 0.1-0.3$ the tension between the value of the Hubble constant as determined from local measurements and as inferred from the temperature anisotropies in the cosmic microwave background can be substantially relaxed~\cite{Riess:2019cxk,Riess:2018byc,Riess:2016jrr}. In the right frame, we show the relationship between the initial mass of the black holes and the temperature of the universe following their evaporation, assuming black hole domination.}
\label{DeltaNeff}
\end{figure}  

After the black holes have evaporated, the universe will be filled with SM radiation, along with any other Hawking radiation products that may have been produced. If there exist additional light states without significant couplings to the SM, so-called ``dark radiation" (DR), these particles will be produced through Hawking radiation and contribute to the radiation density of the universe during BBN and recombination. If the universe is black hole dominated during the era of evaporation, the fraction of the universe's energy density in such particles will be given by the proportion of their degrees-of-freedom, $g_{{\rm DR},H}/g_{\star, H}$. As the universe expands and cools, the energy density in dark radiation is diluted by four powers of the scale factor, 
\be
\label{DRratio}
\frac{ \rho_{\rm DR}(T_{\rm EQ}) }{\rho_{\rm DR}(T_{\rm RH})} = \left(\frac{ a_{\rm RH}}{a_{\rm EQ} }\right)^4.
\ee
However, the energy density in the SM radiation bath is additionally diluted by a series of entropy dumps that occur when SM radiation temperature falls below
the mass of a particle species, which transfers its entropy to the remaining radiation bath and increases the latter's temperature. To evaluate the impact of these transfers, we apply entropy conservation:
\begin{eqnarray}
\label{entropy}
 (a^3 s)_{\rm RH} &=&(a^3 s)_{\rm EQ} \,\,\, \Longrightarrow \,\,\, a^3_{\rm RH} \, g_{\star,S} (T_{\rm RH}) \, T^3_{\rm RH} = a^3_{\rm EQ} \, g_{\star,S} (T_{\rm EQ}) \, T^3_{\rm EQ},
\end{eqnarray}
where $g_{\star, S}$ is the effective number of relativistic degrees of freedom in entropy
\be
g_{\star S}(T) = \sum_{B} g_B \left( \frac{T_B}{T} \right)^3 +  \frac{7}{8}\sum_{F} g_F \left( \frac{T_F}{T} \right)^3~,
\ee
where the sum is over all bosons/fermions $B/F$ with
temperatures $T_{B/F}$ and spin states  $g_{B/F}$. 
Although $g_{\star, S}=g_{\star}$ at high temperatures, this is not the case at matter-radiation equality when $T_{\rm EQ} \simeq 0.75$ eV 
and we have
\be
g_{\star S}(T_{\rm EQ})=2+2  N_{\nu} \left(\frac{7}{8}\right) \left(\frac{4}{11}\right) \approx 3.94~~, ~~~
g_{\star}(T_{\rm EQ})=2+2N_{\nu} \left(\frac{7}{8}\right)  \left(\frac{4}{11}\right)^{4/3} \approx 3.38~,~~
\ee
where $N_\nu \simeq 3.046$ is the effective number of SM neutrinos and we have used Eqs.~(\ref{gstar}) and (\ref{entropy}). Thus, entropy
conservation yields 
\be
\frac{ T_{\rm EQ} }{ T_{\rm RH} }= \left(  \frac{  a_{\rm RH} }{  a_{\rm EQ} }\right) \left(  \frac{  g_{\star, S}(T_{\rm RH}) }{ g_{\star, S}(T_{\rm EQ}) } \right)^{1/3},
\ee
 and the energy density in SM radiation is diluted by the following factor:
\begin{eqnarray}
\label{SMRatio}
 \frac{\rho_{R}(T_{\rm EQ})}{\rho_{R}(T_{\rm RH})} = \bigg(\frac{a_{\rm RH}}{a_{\rm EQ}}\bigg)^4 \, \bigg(\frac{g_{\star}(T_{\rm EQ})}{g_{\star} (T_{\rm RH})}\bigg) \, \bigg(\frac{g_{\star, S}(T_{\rm RH})}{g_{\star, S} (T_{\rm EQ})}\bigg)^{4/3} = \bigg(\frac{a_{\rm RH}}{a_{\rm EQ}}\bigg)^4 \,     \bigg(\frac{g_{\star}(T_{\rm EQ}) \,\, g_{\star, S}(T_{\rm RH})^{1/3}}{g_{\star, S} (T_{\rm EQ})^{4/3}}\bigg). 
\end{eqnarray}
Using Eqs.~(\ref{DRratio}) and (\ref{SMRatio}), the dark and SM radiation density ratio at matter-radiation equality becomes
\begin{equation}
\frac{\rho_{\rm DR}(T_{\rm EQ})}{\rho_{R}(T_{\rm EQ})}  = \bigg(\frac{g_{{\rm DR},H}}{g_{\star, H}}\bigg) \,  \bigg(\frac{g_{\star, S} (T_{\rm EQ})^{4/3}}{g_{\star}(T_{\rm EQ}) \,\, g_{\star, S}(T_{\rm RH})^{1/3}}   \bigg),
\end{equation}
which is related to the effective number of neutrino species via
\be
\Delta N_{\rm eff} =  \frac{\rho_{\rm DR}(T_{\rm EQ})}{\rho_{R}(T_{\rm EQ})} \bigg[N_{\nu} + \frac{8}{7} \bigg(\frac{11}{4}\bigg)^{4/3}\bigg] 
= \bigg(\frac{g_{{\rm DR},H}}{g_{\star, H}}\bigg) \bigg(\frac{g_{\star S}(T_{\rm EQ})}{g_{\star S}(T_{\rm RH})}\bigg)^{1/3} \bigg(\frac{g_{\star S}(T_{\rm EQ})}{g_{\star}(T_{\rm EQ})}\bigg) \bigg[N_{\nu}+\frac{8}{7}\bigg(\frac{11}{4}\bigg)^{4/3}\bigg].
\ee
For high reheat temperatures, we thus have 
\be
\label{neffmain}
\Delta N_{\rm eff}  &\approx& 0.10 \, \bigg(\frac{g_{{\rm DR},H}}{4}\bigg) \bigg(\frac{106}{g_{\star}(T_{\rm RH})}\bigg)^{1/3}, 
\label{darkrad}
\ee
which is one of our main results. Using Eq.~(\ref{neffmain}) we see that, unlike relativistic thermal relics in equilibrium with the SM radiation bath, 
$\Delta N_{\rm eff} \lesssim 0.2$ for any individual decoupled species produced via Hawking radiation, including Dirac fermions. 
This conclusion holds even for low values of $T_{\rm RH} \ll 100$ GeV because the BH ``branching fraction" into
 dark radiation scales as  $g_{{\rm DR}, H}(T_{\rm BH})/ 
g_{\star, H}(T_{\rm BH})$ and $T_{\rm BH} \gg 100$ GeV is always satisfied for BH masses that fully evaporate before BBN ($\tau \ll$ sec), so the
relative DR contribution is always diluted by a factor of $g_{\star, H}(T_{\rm BH}) \simeq 108$.

We summarize these results in Fig.~\ref{DeltaNeff}, where we plot $\Delta N_{\rm eff}$ as a function of the temperature of the universe after the black holes have evaporated, $T_{\rm RH}$. For a light and decoupled Dirac fermion, massive vector, or real scalar, we predict a contribution of $\Delta N_{\rm eff} \simeq 0.03-0.2$, below current constraints from measurements of the CMB and baryon acoustic oscillations (BAO)~\cite{Aghanim:2018eyx}, but within the projected reach of stage IV CMB experiments~\cite{Abazajian:2016yjj,Baumann:2017gkg,Hanany:2019lle}.


If black holes did not dominate the energy density of the early universe, the contribution to $\Delta N_{\rm eff}$ will be reduced accordingly. In the limit in which $\rho_{R} \gg \rho_{\rm BH}$ is maintained throughout the early universe, we arrive at:
\begin{eqnarray}
\Delta N_{\rm eff} \approx 3.5 \times 10^{-5} \times \bigg(\frac{g_{{\rm DR},H}}{4}\bigg) \bigg(\frac{106}{g_{\star}(T_{\tau})}\bigg)^{1/3} \, \bigg(\frac{f_i (10^{10} \, {\rm GeV})}{10^{-15}}\bigg) \bigg(\frac{M_i}{10^{8}\,{\rm g}}\bigg)^{3/2},
\label{darkrad2}
\end{eqnarray}
where $f_i(10^{10} \, {\rm GeV})$ is the BH density fraction $\rho_{\rm BH}/\rho_{\rm R}$ at a temperature of $10^{10}$ GeV and 
$T_\tau$ is the SM temperature at the time of evaporation in Eq.~(\ref{tevap}). In general, we expect Hawking evaporation to generate a sizable contribution to $N_{\rm eff}$ only if the early universe included a black hole dominated era or if there exist a large number of light, decoupled species ($g_{{\rm DR}, H} \gg 1$), as discussed in Sec.~\ref{LHS}.

Up to this point, we have treated the particles that make up the dark radiation as if they were massless. In order for these Hawking radiation products to contribute towards dark radiation, their average kinetic energy evaluated at the time of matter-radiation equality must approximately exceed their mass, $\langle E_{\rm DR} \rangle \gsim m_{\rm DR}$. Under the simplifying assumption that all of the Hawking radiation is emitted at the initial temperature of the black hole, the average energy of these particles is given by:
\begin{eqnarray}
\langle E_{\rm DR} \rangle\bigg|_{\rm EQ} \sim \alpha \,T_{{\rm BH}, i} \times \frac{T_{\rm EQ}}{T_{\rm RH}} 
\bigg(\frac{g_{\star}(T_{\rm EQ})}{g_{\star}(T_{\rm RH})}\bigg)^{1/3}  \sim 3.9 \, {\rm MeV}  \, \bigg(\frac{\alpha}{3.15}\bigg) \bigg(\frac{M_i}{10^8\,{\rm g}}\bigg)^{1/2} \bigg(\frac{108}{g_{\star,H}(T_{\rm BH})} \bigg)^{1/2} \bigg(\frac{14}{g_{\star}(T_{\rm RH})}\bigg)^{1/12}, \nonumber
\end{eqnarray}
where $\alpha \approx 2.7$ (3.15) for bosonic (fermionic) dark radiation. Numerically integrating the deposition of Hawking radiation over the lifetime of the black hole, we arrive at a slightly higher value, a factor of 1.4 larger than that given in the above expression. We thus conclude that in order to contribute towards dark radiation (as opposed to dark matter) at matter-radiation-equality, Hawking evaporation products must be lighter than $\sim 5.5 \, {\rm MeV} \times (M_i/10^8 \, {\rm g})^{1/2}$.

\begin{figure}[t]
\includegraphics[scale=0.54]{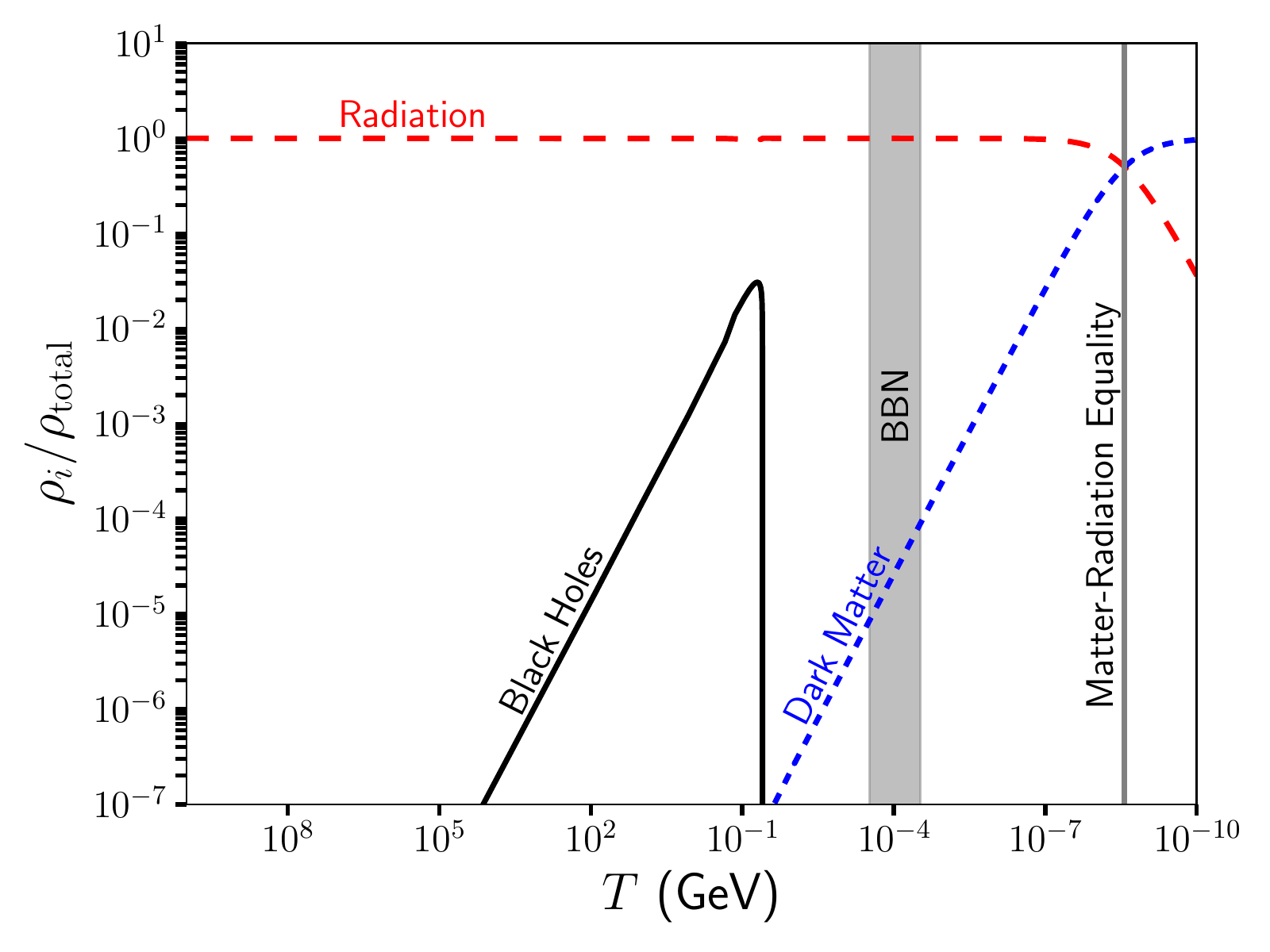} 
\includegraphics[scale=0.54]{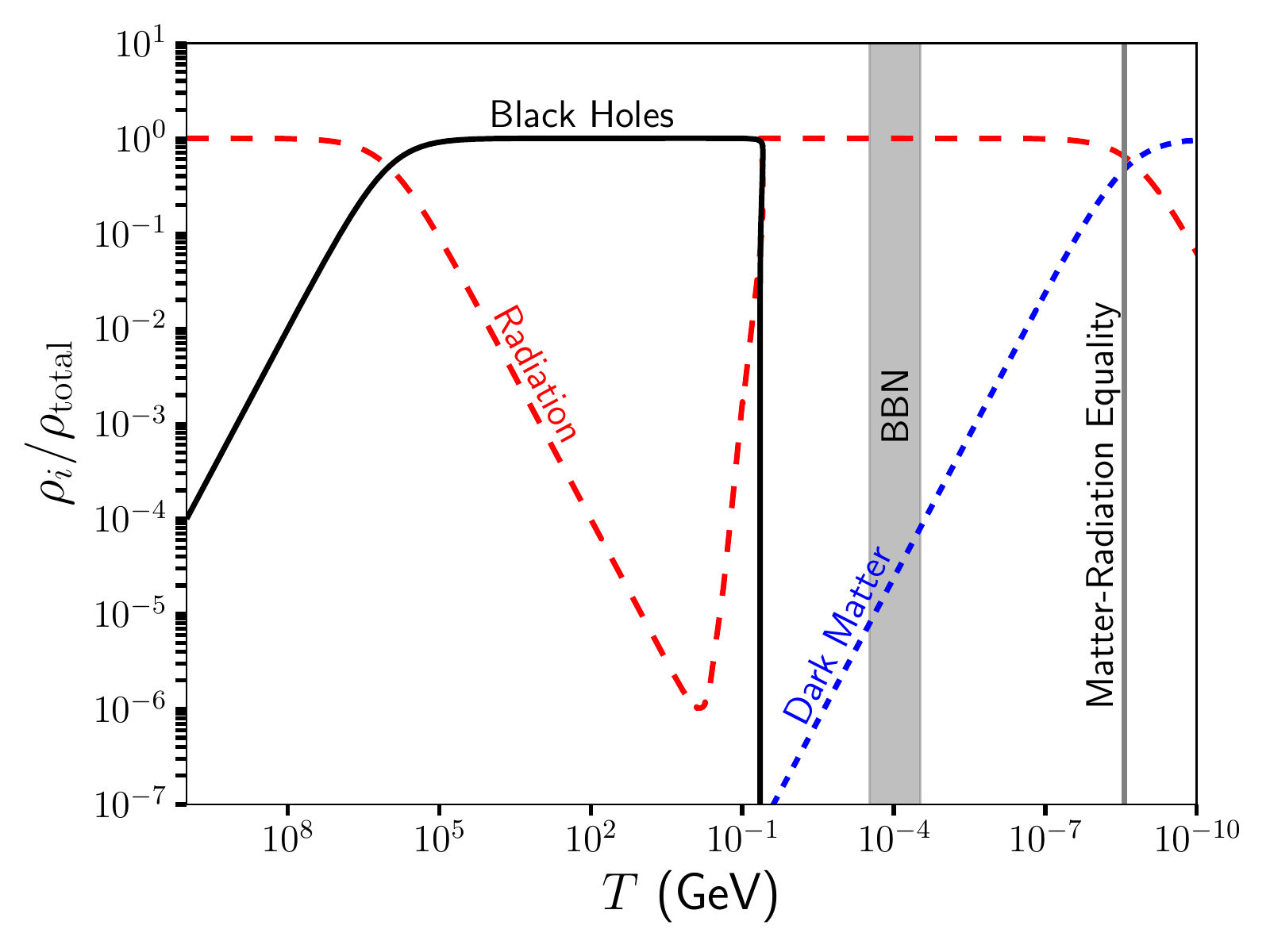} 
\caption{Two examples of the evolution of the energy density in SM radiation, primordial black holes, and dark matter, which is produced through Hawking evaporation. In each frame, we have adopted an initial black hole mass of $M_{i}=10^8$ g. In the left frame, we show results for the case of an initial black hole density of $f_i =8\times 10^{-14}$ at $T_i=10^{10}$ GeV, and a dark matter particle mass of $m_{\rm DM}=10^9$ GeV. In the right frame, the black holes come to dominate the energy density of the universe, and we have chosen $m_{\rm DM}=6\times 10^{10}$ GeV.}
\label{evolution}
\end{figure}

\section{Superheavy Dark Matter as the Products of Hawking Evaporation}

If there exist heavy, stable, decoupled particles, they will also be produced through Hawking radiation and will contribute to the abundance of dark matter (see, for example, Refs.~\cite{Lennon:2017tqq,Morrison:2018xla,Fujita:2014hha,Allahverdi:2017sks}). If the early universe included an era of black hole domination, the evaporation of the black holes leads to the following abundance of dark matter:
\begin{eqnarray}
\Omega_{\rm DM} h^2 \approx 0.1 \, \bigg(\frac{g_{{\rm DM}, H}}{4}\bigg) \, \bigg(\frac{6\times 10^{10} \, {\rm GeV}}{m_{\rm DM}}\bigg) \, \bigg(\frac{10^8 \, {\rm g}}{M_{i}}\bigg)^{5/2}.
\end{eqnarray}
From Eq.~(\ref{domination-fraction}), if the initial density ratio satisfies $\rho_{{\rm BH},i}/\rho_{{R},i} \lesssim 4\times 10^{-12} \times (10^{10} \,{\rm GeV}/T_i)(10^8 {\rm g}/M_i)^{3/2}$,
 the BH population never dominates the total energy density of the universe. In this case, the evaporation of the black holes leads to the following abundance of dark matter:
\begin{eqnarray}
\Omega_{\rm DM} h^2 \approx 0.1 \, \bigg(\frac{f_i (10^{10} \, {\rm GeV})}{8 \times 10^{-14}}\bigg) \,  \bigg(\frac{g_{{\rm DM},H}}{4}\bigg) \, \bigg(\frac{10^9 \, {\rm GeV}}{m_{\rm DM}}\bigg) \, \bigg(\frac{10^8 \, {\rm g}}{M_{i}}\bigg).
\end{eqnarray}

In Fig.~\ref{evolution} we show two examples of the evolution of energy densities in radiation, black holes, and dark matter (as produced via Hawking radiation). In the left frame, we start our Boltzmann code (at $T=10^{10}$ GeV) with only a small abundance of black holes, $f_i \equiv \rho_{R}/\rho_{\rm BH} = 10^{-14}$. In this case, the universe never becomes black hole dominated, and the Hawking evaporation of Dirac fermions with a mass of $10^9$ GeV make up the measured abundance of dark matter. In the right frame, we instead consider a case in which the early universe becomes black hole dominated, for which a heavier dark matter candidate is required. In each frame, we consider black holes with an initial mass of $M_i = 10^8$ g.

\begin{figure}[t]
\includegraphics[scale=0.6]{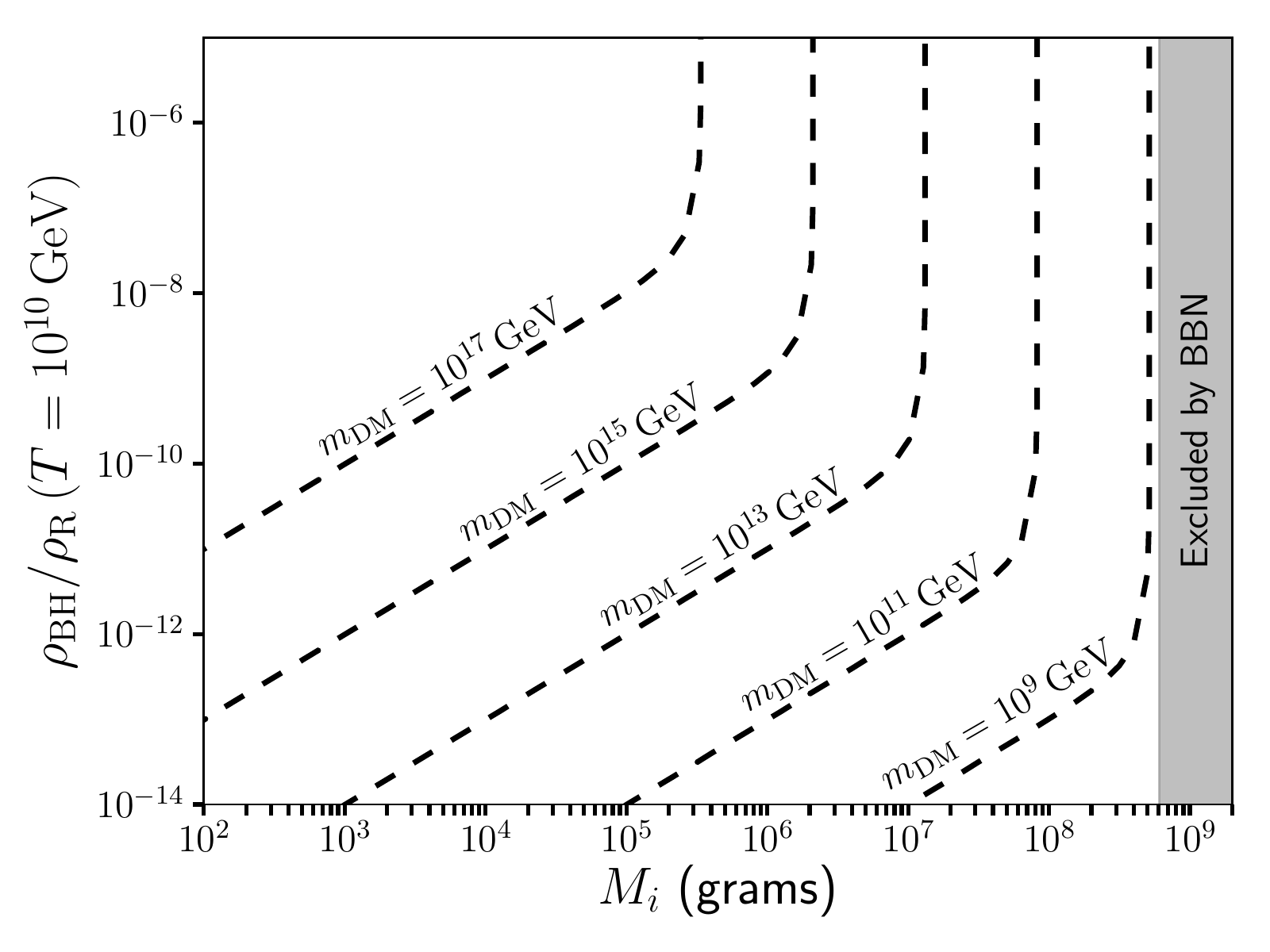} 
\caption{The values of the black hole energy fraction $\rho_{\rm BH}/\rho_{R}$ (evaluated at $T=10^{10}$ GeV) and initial black hole mass $M_i$ that lead to $\Omega_{\rm DM}h^2 \simeq 0.1$ for the case of dark matter in the form of a Dirac fermion with negligible couplings to the SM. Throughout the upper right portion of this plane, the early universe included a period in which black holes dominated the total energy density, and thus the results do not depend on the initial value of $\rho_{\rm BH}/\rho_{R}$.
}
\label{plane}
\end{figure}

In Fig.~\ref{plane}, we show the values of $\rho_{\rm BH}/\rho_{R}$ (evaluated at $T=10^{10}$ GeV) and the initial black hole mass that lead to $\Omega_{\rm DM}h^2 \simeq 0.1$, for the case of dark matter in the form of a Dirac fermion with negligible couplings to the SM. In the upper right (lower left) portion of this plane, the early universe included (did not include) a period in which black holes dominated the total energy density.

Next, we consider the case in which each black hole ends the process of evaporation leaving behind a remnant of mass $M_{\rm remnant} = \eta    M_{\rm Pl}$~\cite{MacGibbon:1987my,Carr:1994ar,Barrow:1992hq}. In a scenario in which black holes never dominate the energy density of the universe, the black holes lead to the following abundance of such remnants:
\begin{eqnarray}
\Omega_{\rm remnant} h^2 \approx 0.1 \times \bigg(\frac{\eta}{1}\bigg) \,  \bigg(\frac{f_i (10^{10} \, {\rm GeV})}{1.5 \times 10^{-6}}\bigg) \,  \bigg(\frac{10^8 \, {\rm g}}{M_{i}}\bigg).
\end{eqnarray}

In the opposite limit, in which there was a black hole dominated era, the evaporation of the black holes leads to the following abundance of Planck mass remnants:
\begin{eqnarray}
\Omega_{\rm remnant} h^2 \approx 0.1 \times  \bigg(\frac{\eta}{1}\bigg) \,   \bigg(\frac{6 \times 10^5 \, {\rm g}}{M_{i}}\bigg)^{5/2}.
\end{eqnarray}

Dark matter in the form of a superheavy, gravitationally interacting state would be very challenging to observe or otherwise test. It has been proposed, however, that an array of quantum-limited impulse sensors could be used to detect gravitationally particles with a Planck-scale mass~\cite{Carney:2019pza}.


\section{Primordial Black Holes in the Presence of Large Hidden Sectors}
\label{LHS}

So far in this study, we have assumed that the black holes evaporate mostly into SM particles, possibly along with a small number of states that act as either dark radiation or dark matter. It seems highly plausible, however, that the SM describes only a small fraction of the degrees-of-freedom that constitute the universe's total particle content. If a large number of other particle degrees-of-freedom exist, black holes will evaporate more rapidly, producing the full array of particles that are kinematically accessible (all of those with masses below $\sim T_{\rm BH}$), independently of their couplings or other characteristics.  

As a first case, we will consider a scenario in which there exist a large number of degrees-of-freedom associated with light particles with negligible couplings to the SM. Such a situation is motivated, for example, within the context of the string axiverse, in which a large number of light and feebly-coupled scalars are predicted~\cite{Svrcek:2006hf,Svrcek:2006yi,Arvanitaki:2009fg,Fox:2004kb}. If the early universe experienced a black hole dominated era, it follows from Eq.~(\ref{darkrad}) that this would lead to $\Delta N_{\rm eff} \sim (0.04-0.08) \times N_{\rm axion}$, where $N_{\rm axion}$ is the number of axions that exist (see also Fig.~\ref{DeltaNeff}). Given the current constraint of $\Delta N_{\rm eff} \lsim 0.28$~\cite{Aghanim:2018eyx}, this indicates that $N_{\rm axion} \lsim 7$, regardless of $T_{\rm RH}$. Thus the existence of a black hole dominated era appears to be inconsistent with the existence of a large axiverse (see also Ref.~\cite{Lennon:2017tqq}). 

As a second example, consider the Minimal Supersymmetric Standard Model (MSSM). In this case, approximately half of the Hawking evaporation products will be superpartners for all black holes with a temperature greater than the characteristic scale of superpartner masses, $T_{\rm BH} \gsim M_{\rm SUSY}$, corresponding to $M_{\rm BH} \lsim 10^{10}$ g $\times ({\rm TeV}/M_{\rm SUSY})$. If $R$-parity is conserved, all such superpartners will decay to the lightest supersymmetric particle, producing a potentially large relic abundance. If the lightest superpartner is weakly interacting (such as a neutralino), it may or may not reach equilibrium with the SM bath, depending on the temperature at the time of evaporation. For $M_i \gsim 10^{-2}$ g $\times ({\rm TeV}/M_{\rm SUSY})^{2/3}$, Hawking evaporation will finalize at a temperature below that of neutralino freeze-out, leading to a large relic abundance. If the early universe included a black hole dominated era, such a scenario is strongly excluded (see also Refs.~\cite{Green:1997sz,Green:1999yh,Lemoine:2000sq,Khlopov:2004tn}). To obtain an abundance of superpartners that is equal to the measured density of dark matter, we would require an initial black hole abundance (at $T=10^{10} \, {\rm GeV}$) of only $\rho_{\rm BH}/\rho_{R} \sim 10^{-20} \, ({\rm TeV}/M_{\rm SUSY}) (10^8\,{\rm g}/M_i)$.

\section{Summary and Conclusions}

\begin{figure}[b!]
\includegraphics[scale=0.5]{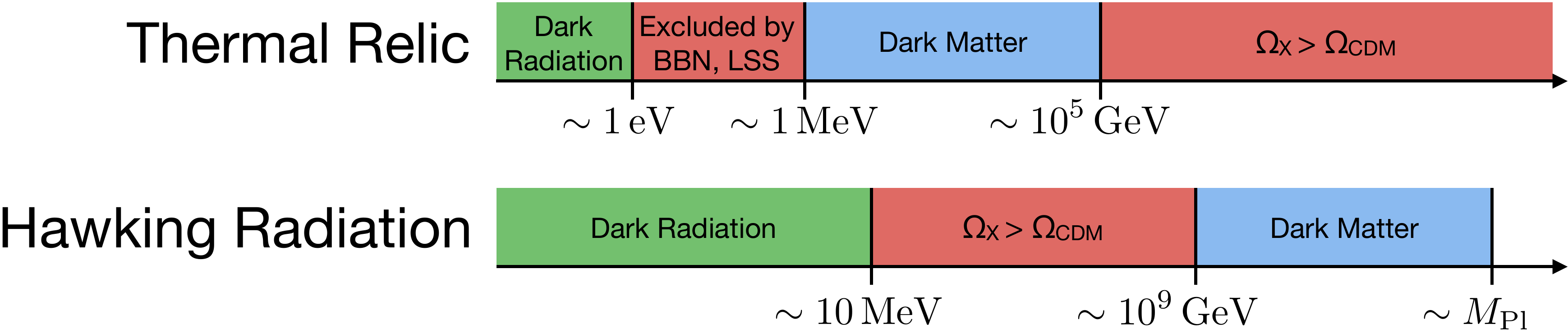} 
\caption{In contrast to thermal relics, Hawking radiation in a black hole dominated scenario can produce dark radiation in the form of particles with masses up to $m_{\rm DR} \sim 5.5 \, {\rm MeV} \times (M_i/10^8 \, {\rm g})^{1/2}$, as well as dark matter for particles with masses between $m_{\rm DM} \sim 10^{11} \, {\rm GeV} \times (10^8 \, {\rm g}/M_i)^{5/2}$ and the Planck scale. For thermal relics, masses below $\lesssim$ MeV spoil early universe cosmology \cite{nollett:2013pwa} and masses above $\gtrsim 10^{5}$ GeV overclose the universe in perturbative, unitary theories \cite{griest:1989wd}.
}
\label{skematic}
\end{figure}

If even a small abundance of black holes were present in the early universe, this population would evolve to constitute a larger fraction of the total energy density, up to the point at which they evaporate or until matter-radiation equality. From this perspective, it is natural to consider scenarios in which the early universe included an era of black hole domination. To avoid altering the light element abundances, such black holes must evaporate prior to BBN, corresponding to initial masses less than $M_i \lsim 6\times 10^8$ g.

Unlike most other mechanisms for particle production in the early universe, Hawking evaporation generates particles democratically, producing all particle species including those with with small or negligible couplings to the SM. From this perspective, black holes provide a well-motivated mechanism to produce both dark radiation and dark matter. If the early universe included a black hole dominated era, then Hawking radiation will contribute to dark radiation at a level $\Delta N_{\rm eff} \sim 0.03-0.2$ for each light and decoupled species, depending on their spin and on the initial black hole mass. This range is well suited to relax the tension between late and early-time Hubble determinations, and is within the reach of upcoming stage IV CMB experiments. The dark matter could also originate as Hawking radiation in a black hole dominated early universe, although such dark matter candidates must be very heavy to avoid exceeding the measured abundance, $m_{\rm DM} \sim 10^{11} \, {\rm GeV} \times (10^8 \, {\rm g}/M_i)^{5/2}$.

In Fig.~\ref{skematic} we summarize some of key results of this analysis and contrast them with those found in the case of dark radiation or dark matter that originates as a thermal relic of the early universe. Whereas thermal relics can constitute dark radiation only if lighter than $\sim$ eV, Hawking radiation in a black hole dominated scenario can produce dark radiation in the form of particles with masses up to $m_{\rm DR} \lsim 5.5 \, {\rm MeV} \times (M_i/10^8 \, {\rm g})^{1/2}$. Similarly, whereas thermal relics can make up the dark matter without unacceptably altering the light element abundances and large scale structure of the universe only for $m \gsim \, {\rm MeV}$, Hawking radiation can produce an acceptable abundance of dark matter particles with masses between $m_{\rm DM} \sim 10^{11} \, {\rm GeV} \times (10^8 \, {\rm g}/M_i)^{5/2}$ and the Planck scale.

\begin{acknowledgments}  
This manuscript has been authored by Fermi Research Alliance, LLC under Contract No. DE-AC02-07CH11359 with the U.S. Department of Energy, Office of High Energy Physics. 
\end{acknowledgments}

\bibliography{PBH}

\appendix
\section{Black Hole Mergers in the Early Universe}
\label{mergers}

The rate at which black holes are captured into binary systems is given by $\Gamma_{\rm bc} = n_{\rm BH} \sigma_{\rm bc} v$. For two black of holes of the same mass, this cross section is given by~\cite{Bird:2016dcv}:
\beq
\sigma_{\rm bc} = \pi \pL \frac{85 \pi}3 \pR^{2/7} r_{\rm Schw}^2 v^{-18/7} \simeq 45 \pL \frac{M_{\rm BH}}{M_{\rm Pl}^2} \pR^2 v^{-18/7},
\eeq
where $r_{\rm Schw}$ is the Schwarzschild radius of the black holes. The ratio of the binary capture rate to that of Hubble expansion is thus given by:
\begin{eqnarray}
\frac{\Gamma_{\rm bc}}{H} &\approx& \frac{45 \sqrt{3} \rho_{\rm BH} M_{\rm BH} v^{-11/7}}{\sqrt{8\pi} \rho^{1/2}_{\rm tot} M^3_{\rm Pl}} \\
& \approx & 0.02 \times \bigg(\frac{M_{\rm BH}}{10^8 \, {\rm g}}\bigg) \, \bigg(\frac{T_{\rm eff}}{10^7 \, {\rm GeV}}\bigg)^2 \, \bigg(\frac{v}{10^{-5}}\bigg)^{-11/7} \,\bigg(\frac{\rho_{\rm BH}}{\rho_{\rm tot}}\bigg),
\end{eqnarray}
where $T_{\rm eff}$ is defined such that $\rho_{\rm tot} \equiv \pi^2 g_{\star}(T_{\rm eff}) T^4_{\rm eff}/30$. We conclude that even if the energy density of the early universe were dominated by black holes, the rate for binary capture exceeds the rate of Hubble expansion only at very early times, $T_{\rm eff} \gsim 10^8 \, {\rm GeV} \times (10^8 \, {\rm g}/M_{\rm BH})^{1/2} \, (v/10^{-5})^{11/14}$.

Furthermore, even if primordial black holes form binaries efficiently in the early universe, it is not clear that they will merge before evaporating. Assuming that gravitational wave emission dominates the process by which black hole binaries loses energy, the inspiral time is given by~\cite{Peters:1964zz}: 
\beq
t_{\rm insp} = \frac{5 a_0^4}{512 G^3 M_{\rm BH}^3},
\eeq
where $a_0$ is the initial separation of the inspiring black holes. The ratio of the inspiral time to the evaporation time [see Eq.~(\ref{tevap})] is then given by:
\beq
\frac{t_{\rm insp}}{\tau} \simeq 10 \times \zeta^4 \pL \frac{10^8\gev}{T_{\rm eff}} \pR^8 \pL \frac{10^8 \g}{M_{\rm BH}} \pR^6,
\eeq
where $\zeta \equiv a_0 H_{\rm bc}$ and $T_{\rm eff}$ is again defined such that $\rho_{\rm tot} \equiv \pi^2 g_{\star}(T_{\rm eff}) T^4_{\rm eff}/30$, but in this case evaluated at the time of binary capture. Since $\zeta \sim \cO(0.1)$, we expect the black holes to merge before evaporating ($t_{\rm insp} \lsim t_\tau$) only at very early times, $T_{\rm bc} \gsim 4\times 10^7 \, {\rm GeV} \times (\zeta/0.1)^{1/2} (10^8 \, {\rm g}/M_{\rm BH})^{3/4}$. We do note, however, that in the early universe, when the ambient density is large, it is possible that more efficient mechanisms of angular momentum transfer were active, potentially leading to shorter inspiral times.

To summarize this section, if the universe were black hole dominated at very early times, corresponding to effective temperatures greater than $\sim 10^{8}$ GeV $\times (10^8 \, {\rm g}/M_{\rm BH})^{3/4}$, a substantial fraction of the black holes may have undergone mergers. With this in mind, one should interpret the initial black hole mass, $M_i$, as used throughout this paper to denote the mass of the black holes after the processes of binary capture and inspiral have become inefficient.

\section{Bondi-Hoyle Accretion}
\label{accretion}

A black hole in a bath of radiation will undergo Bondi-Hoyle accretion, gaining mass at the following rate~\cite{Bondi:1952ni}:
\beq \label{massacc1}
\frac{dM_{\rm BH}}{dt}\bigg|_{\rm Accretion} = \frac{4\pi \lambda M_{\rm BH}^2 \rho_{R}}{M_{\rm Pl}^4(1+c_s^2)^{3/2}},   
\eeq
where $\lambda$ is an $\cO(1)$ constant and $c_s \simeq 1/\sqrt{3}$ is the sound speed in the radiation bath. Combining this with the rate of mass loss from Hawking evaporation [see Eq.~(\ref{rate})], we can write the total rate of change as follows:
\beq \label{massacc2}
\frac{dM_{\rm BH}}{dt} =  \frac{\pi \mathcal G g_{*,H}(T_{\rm BH}) T_{\rm BH}^2}{480} \bL \frac{ \lambda g_*(T_{R})}{\mathcal G g_{*,H}(T_{\rm BH}) (1+c_s^2)^{3/2}} \pL \frac{T_{R}}{T_{\rm BH}} \pR^4 -  1 \bR.
\eeq
Since $\lambda g_*/\mathcal G g_{*,H} (1+c_s^2)^{3/2}$ is an order one quantity, we conclude that a black hole will generally gain mass when $T_{R} \gsim T_{\rm BH}$, and lose mass otherwise. When we compare the fractional accretion rate, $(1/M)dM/dt$, to that of Hubble expansion, we find that accretion plays an important role only when $T_{R} \gsim 10^{12} \, {\rm GeV} \times (10^8 \, {\rm g}/M_{\rm BH})^{1/2}$.

\end{document}